# Controlling the electrostatic Faraday instability using superposed electric fields

Sebastian Dehe[1], Maximilian Hartmann[1], Aditya Bandopadhyay[2], Steffen Hardt[1,*]

[1]*Fachgebiet Nano- und Mikrofluidik, Fachbereich Maschinenbau, TU Darmstadt, 64287 Darmstadt, Germany*

[2]*Department of Mechanical Engineering, Indian Institute of Technology Kharagpur, Kharagpur- 721302, India*

When the interface between a dielectric and a conducting liquid is excited by an oscillatory electric field, electrostatic Faraday waves can be induced. Here, we study the response of the interface to an AC electric field, which is superposed by either a second AC field of different frequency, or by a DC field. An algorithm based on light refraction at the fluidic interface is used to obtain the spatio-temporal response of the Faraday waves, and the critical voltage corresponding to onset of instability, the interfacial oscillation frequency and the dominant wavelengths are determined. The influence of the mixing ratio, which denotes the relative amplitudes of the different components of the driving signal, is analyzed, and the experimental results are compared with theoretical predictions. For AC/AC driving, gradual variations of the mixing ratio can induce a jump of the pattern wavelength, which is a result of the transition from harmonic to a subharmonic oscillation. For AC/DC driving, the interface oscillates either harmonically or subharmonically, and the response wavelength can be tuned continuously by adjusting the admixture of the DC component. For both driving modes, the experiments show good agreement with theory.

The parametric excitation of a liquid layer under oscillatory actuation has been studied since the initial experiments by Faraday in 1831, who observed the instability of a liquid layer deposited on top of a vibrating plate [1]. In the following years, Faraday waves were reported to occur with either half the driving frequency (subharmonic response) or isochronous with the driving frequency (harmonic response) [1–4]. Ultimately, Benjamin and Ursell showed that the inviscid case is well described by a Mathieu equation, permitting both response frequencies [5]. Since these initial works, the Faraday instability has been studied extensively as a model for unstable systems. For example, investigations have focused on viscosity effects [6], the specific modes of the instability [5,7–10], secondary instabilities [11], time averages of the instability patterns [12], amplitude- and phase-resolved measurements of specific patterns [13], as well as hysteresis effects [9,14–16]. Also, the influence of the domain boundary has been studied extensively [8–10,17–19]. Multiple reports of the system response to a driving consisting of more than one frequency component exist, with observations of quasi-patterns that exhibit long-range order but no spatial periodicity [10,20–22] and mode interactions [23–25]. Aside its role as a model system, the Faraday instability has been used for measuring interfacial tensions [6,26] and for creating defined patterns of biological samples, such as cells [27–29].

While the above-mentioned work all refers to instabilities induced by a time-periodic acceleration, wave patterns on a liquid surface or liquid-liquid interface can also be induced by a time-periodic electric field. A resonant response of the interface between a dielectric and a conducting liquid can be induced by an oscillatory electric field, and the governing equations for the inviscid case are similar to those of the mechanically actuated Faraday instability [30]. Since the Maxwell stress at the interface is proportional to the square of the electric field, the lowest accessible frequency of a single-frequency harmonic actuation is isochronous with the driving field. While the electrostatic forcing of Faraday waves has attracted less attention than mechanical forcing, electric-field induced instabilities were studied in the context of surface tension measurements [31] and within ozone generators [32–34]. Also, the response of a liquid film to direct current (DC) and alternating current (AC) actuation was studied in the context of pillaring instabilities [35], as well as the instability of liquid interfaces under AC forcing [36–39]. Recently, a theoretical description of the electrostatic Faraday instability including viscosity effects was provided for the interface between a perfect dielectric and a perfect conductor, both for single-frequency and multiple-frequency actuation [40]. Subsequently, the theory was extended to account for leaky-dielectric liquids, and experiments showed that the critical voltage corresponds to theoretical predictions for single-frequency actuation, both with and without an additional constant field component [41]. Recently, our own experiments demonstrated that the resulting wavelengths under single-frequency actuation without an additional DC offset agree with the theoretical predictions [42]. Also, the resulting instability patterns showed similarities to patterns obtained for mechanical forcing, further highlighting the analogies between electric and mechanical forcing.

---

[*] Corresponding author. hardt@nmf.tu-darmstadt.de

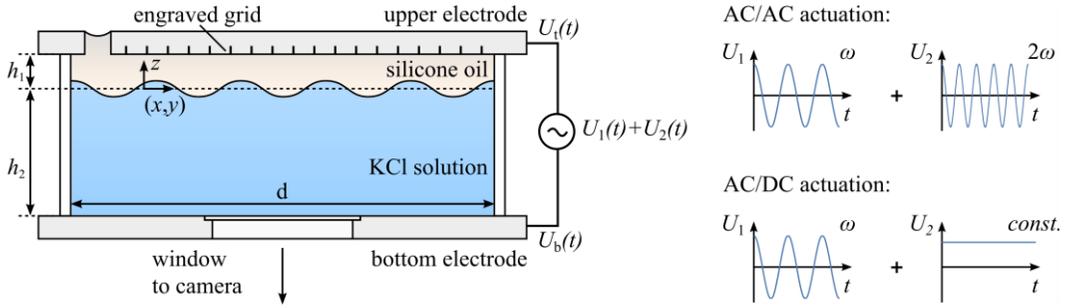

FIG.1. Schematic of the experimental setup for studying the electrostatic Faraday instability. A circular container is filled with a dielectric and a conducting liquid. A voltage is applied between the upper and lower electrode, and induces an instability at the liquid-liquid interface. To study complex actuation, the driving signal consists of an AC component that is superposed by either a second AC component or a DC component. A laser-engraved grid at the upper electrode is imaged through the liquid-liquid interface, and distorted due to refraction. From the distorted image, the local thickness $h(x, y)$ of the upper liquid layer can be reconstructed. The undeformed interface is indicated by the dashed line.

However, electric forcing offers some opportunities not accessible with mechanical forcing. Notably, while mechanical forcing does not allow imposing localized inertial forces on a liquid film without introducing major perturbations to the system, localized electric forces can be imposed using a pin electrode. Further, applying a time-independent acceleration is highly non-trivial, while a steady-state electric force is easily feasible. In this letter, we explore the unique possibilities of electric forcing. We present experimental results for the electrostatic Faraday instability beyond single-frequency actuation. The driving electric fields consist of an oscillatory component superposed either by a second oscillatory component (AC/AC actuation) or by a constant electric field (AC/DC actuation). Changing the relative magnitude of the different frequency components in AC/AC actuation allows to trigger either a harmonic or subharmonic response, accompanied by a jump of the instability wavelength. In AC/DC actuation, the pattern wavelength can be tuned continuously within a certain interval by increasing the relative magnitude of the DC field component, thus enabling an additional degree of freedom compared to mechanical actuation.

*Experimental details.* Figure 1(a) shows a schematic of the experimental setup, where we use light refraction at the liquid-liquid interface to reconstruct the wave patterns. It is similar to the setup we used to study the electrostatic Faraday instability under single-frequency forcing [42]. A glass cylinder (height $h_{cyl} = 35$ mm, diameter $d = 125$ mm) is fixed between two circular stainless-steel electrodes and serves as the fluidic domain. Gaskets (EPDM, APSOparts, Germany) between the cylinder and the electrodes ensure a leakage-free fit, and the system is fixed via a plastic holder (not shown). Optical access is provided via a rectangular cut-out (46 mm x 46 mm) in the bottom electrode, in which a glass window is glued using UV adhesive (NOA68, Thorlabs, Germany). The liquids are added via a filling port (diameter 10 mm) close to the edge of the upper electrode. Also, the upper electrode is structured with a laser-engraved grid (line distance 0.5 mm), which we image through the window using a high-speed camera (FASTCAM Mini AX, Photron, Japan) and a macro objective (SWM VR ED IF Micro 1:1, Nikon, Japan). The resulting observation region at the center of the cell has a side length of approximately 35 mm.

For the results presented in this letter, we used a silicone oil (kinematic viscosity $\nu_1 = 0.65$ cSt, layer thickness $h_2 = 5$ mm, Silikonöl AK 0.65, Silikon Profis, Germany) as the dielectric liquid, and a mixture of de-ionized water and glycerol with additional KCl (60 wt% glycerol, kinematic viscosity $\nu_2 = 8.14$ cSt, KCl concentration $c_{KCl} = 1$ mM, layer thickness $h_2 = 30$ mm) as the conducting liquid. Prior to filling the container, we degas both liquids to avoid gas bubbles from forming. The liquids are introduced using pipettes to avoid splashing and deposition of the conducting liquid at the upper electrode. To prevent dust from entering the container, the filling port is covered prior to experiments. After filling, the optical components are aligned and focused to provide a distortion-free image of the grid at the upper electrode.

The Faraday waves are induced by providing a voltage between the plate electrodes using a function generator (HMF2550, HAMEG Instruments, Germany), generating an output signal between $\pm 10$ V peak to peak. Subsequently, a high-voltage amplifier (HA3B3-S, hivolt.de, Germany) increases the voltage amplitude, providing potentials at both electrodes with opposite signs ($U_t = -U_b = 0.5\, U(t)$). Here, $U(t)$ denotes the voltage between both electrodes. Thereby, almost arbitrary functional forms of the voltage can be created, with a maximum of $\pm 6000$ V peak-to-peak, and a sufficiently large frequency range. For the AC/AC actuation, voltages of the form

$$U(t) = U_0(\chi \cos(\omega t) + (1-\chi)\cos(2\omega t)) \qquad (1)$$

are applied, where $U_0$ denotes the amplitude, $\omega$ the base driving frequency, and $\chi$ the mixing ratio of the two signals, which varies between 0 and 1. We chose the superposition of two oscillatory signals with a ratio of the frequencies of 1/2, which has been investigated in context with multi-frequency forcing of mechanical Faraday waves [43–45]. Then, the theoretical description of Bandopadhyay and Hardt [40] allows us to obtain the critical voltage and wavelength of the instability, as we will outline below. For AC/DC actuation, the voltage is applied as

$$U(t) = U_0(\chi + (1-\chi)\cos(\omega t)). \qquad (2)$$

Thus, for a mixing ratio of $\chi = 0$, the voltage oscillates around zero. For $\chi = 1$, the voltage is a DC signal. All voltage signals at the outlet port of the high-voltage amplifier are monitored with an additional digital oscilloscope (DSO 4022, VOLTCRAFT, Germany).

In a typical experiment, we switch on the power source to provide a defined voltage according to Eqs. (1) and (2), exciting the liquid-liquid interface. Several typical responses can be noted, depending on the excitation amplitude. For voltages well below the onset threshold of Faraday waves, the meniscus of the liquid-liquid interface at the sidewall of the container oscillates due to the varying Maxwell stress. As the Maxwell stress is proportional to the square of the electric field, it oscillates with frequency components of up to $4\omega$ for AC/AC actuation and $2\omega$ for AC/DC actuation. These oscillations induce waves that travel inside the container, and show a distinct circular pattern. In the following, they will be referred to as edge waves. Upon increasing the amplitude, Faraday waves start to appear gradually above the instability threshold, with wave amplitudes increasing over time. The interface can oscillate harmonically or subharmonically, depending on the mixing ratio χ. For two-frequency forcing of the form of Eq. (1), the response is characterized as subharmonic (harmonic) in case of oscillations with odd (even) integer multiples of $\omega/2$. Usually, for a given voltage amplitude the wave amplitude saturates due to nonlinear dampening of the instability [46]. Upon further increase of the voltage, from a certain point on the amplitude of the Faraday waves starts increasing indefinitely. Ultimately, the conducting liquid gets in contact with the upper electrode and forms a conducting bridge, triggering current-limited shutoff. If the interface exhibits Faraday waves, we record the deformed grid at the upper side of the electrode with 1000 frames per second, starting 150 s after the forcing was switched on.

The evaluation of experimental data is based on light refraction at the liquid-liquid interface, which was described by Moisy et al. [47], and recently used in our own experiments to determine the spatial structure of the electrostatic Faraday instability under single-frequency forcing [42]. In short, we image the grid at the upper electrode through the liquid-liquid interface. Since both liquids have different indices of refraction (dielectric liquid $n_1 = 1.376$, conducting liquid $n_2 = 1.412$), refraction at the perturbed interface distorts the image of the grid. For the system shown in Fig. 1, the local thickness $h(x, y)$ of the upper layer can be determined when specific conditions are fulfilled. Under the following assumptions, the interface gradient $\nabla h$ is proportional to the displacement field $\Delta \mathbf{x}$, which represents the difference between the image of the original grid and that of the recorded grid: First, the optics has to be paraxial, requiring that the observation field size $L$ is small compared to the distance $H_{\text{cam}}$ between the interface and the objective ($L/H_{\text{cam}} \approx 35 \text{ mm}/350 \text{ mm} \ll 1$). Second, the interface slope, which is characterized by the angle between the surface normal $\mathbf{n}$ and the unit vector in $z$-direction, has to be small. Third, the interface deformation has to be small ($(h_1 - h)/h_1 \approx 0$). Then, we can express the surface gradient as

$$\nabla h = -a_{\text{cal}} \left( \frac{n_1}{(n_1 - n_2)h_1} - \frac{1}{H_{\text{cam}} + h_1} \right) \Delta \mathbf{x}, \tag{3}$$

where $a_{cal}$ is a calibration factor with units of $\text{mm}/\text{px}^{-1}$. After obtaining the interface gradient, the local interface deflection $\Delta h(x, y) = h - h_1$ can be reconstructed using a numerical algorithm, described in detail elsewhere [42,47]. Thereby, we are able to obtain the three-dimensional shape of the interface from our 2D measurements.

In order to extract the dominant pattern wavelengths, we evaluate the instability over two oscillation periods of the base frequency $\omega$, thus recording at least one full oscillation period of both harmonic as well as subharmonic interface oscillations. The dominant pattern wavelengths can be extracted from each image based on a Fourier transform and computing the one-dimensional power spectrum (see Appendix C of Ref. [42]). The power spectrum exhibits one or more peaks, which correspond to wavelengths present in the image. Using the Python toolbox scipy, we locate the peaks in the spectrum, and subsequently fit a Gaussian distribution to the neighborhood of each peak. The mean value of the distribution is the resulting wavelength associated with each peak. In Fig. 2(a), an exemplary dataset for one experiment is shown, in which multiple wavelengths are detected. Here, both harmonic and subharmonic oscillations are present in the system, creating two distinct, superposed wavelengths (see Fig. 2(b,c)). Creating the histogram of wavelengths over all images shows a bimodal distribution of the wavelengths present in one experiment. In the following, we show the standard deviation of each individual wavelength obtained during the evaluation of one experiment as error bars. Also, we verified for each experiment that the system response results from Faraday waves and not from edge waves, which can be identified based on their oscillation frequency, as well as their circular pattern.

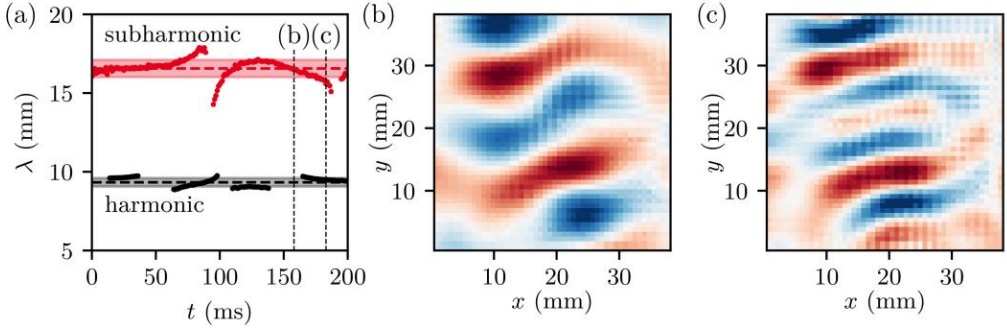

FIG.2. Experimental evaluation of the dominant wavelengths of the instability for AC/AC actuation (actuation parameters: $f = \omega/2\pi = 10$ Hz, $U_0 = 4650$ V, $\chi = 0.8$). (a) Experimentally obtained wavelengths over time, showing a bimodal distribution. Both harmonic and subharmonic oscillations are superposed, which is associated with two distinct wavelengths and response frequencies. The mean values ± standard deviations of the modes are shown as shaded areas. (b,c) Reconstructed interfaces corresponding to the times indicated in (a). The normalized interface deflection is displayed, with positive deflections in red and negative deflections in blue. Here, the two dominant wavelengths are visible.

*Theoretical description.* The theoretical description follows closely our previous work on the electrostatic Faraday instability [40,42], and only the key ideas are repeated here. Essentially, we describe the instability using a domain perturbation approach for the fluid interface, combined with Floquet theory. The mathematical description is based on the assumption that the upper liquid is a perfect dielectric, and the lower liquid a perfect conductor. Furthermore, the upper liquid layer is assumed to be thin compared to the lower layer ($h_1/h_2 \ll 1$), simplifying the resulting equation system. The model accounts for viscous forces in both liquids, capillary effects at the interface (interfacial tension $\sigma = 27.4$ mN m$^{-1}$), and gravity acting on the system. Mathematically, the problem reduces to a generalized eigenvalue problem, which can be solved to yield the marginal stability curve for a given perturbation wavenumber $k$. The minimum of the resulting marginal stability curve in the voltage-wavenumber space determines the critical voltage $U_{\text{crit}}$, above which we expect Faraday waves to occur, as well as the associated most-unstable wavenumber $k_{th}$. We obtain two marginal stability curves, one for the subharmonic and one for the harmonic response of the interface. In experiments, we expect to observe the response with the lower critical voltage $U_{\text{crit}}$, which varies with the mixing ratio $\chi$ for otherwise constant parameters (fluid parameters and base frequency $\omega$). Conversely, changing the mixing ratio allows us to control the response of the interface in terms of oscillation frequency and wavelength.

*AC/AC actuation.* We study the instability induced by the superposition of two AC signals of the form of Eq. (1), where we set the base frequency to $f = \omega/2\pi = 10$ Hz. We conducted experiments for varying mixing ratios $\chi$ with voltage amplitudes $V_0$ starting well below the theoretically predicted critical voltage, as is shown in Fig. 3(a). The critical voltages for the harmonic and subharmonic responses (shown as dotted and dashed lines) have two intersection points (bicritical points), located approximately at $\chi_1 \approx 0.14$ and $\chi_2 \approx 0.77$. For the intermediate mixing ratio range $\chi_1 < \chi < \chi_2$, the subharmonic response has a lower critical voltage, and thus the interface is expected to oscillate subharmonically. Outside of this range, theory predicts harmonic oscillations. As is visible from Fig. 3(a), the experimentally observed critical amplitude of the instability agrees fairly well with theory. The experimentally determined critical voltage shows a local maximum close to the first intersection point $\chi_1$. However, we were unable to observe the second local maximum predicted by theory close to $\chi_2$. For the mixing ratio of $\chi = 0$ (purely harmonic signal with frequency $f = 20\,Hz$), the Faraday waves oscillate with a frequency of $20\,Hz$. In the intermediate mixing ratio range, the interface oscillates with $5\,Hz$ in the subharmonic branch. For signals with $\chi > \chi_2$, the Faraday waves oscillate with $10\,Hz$.

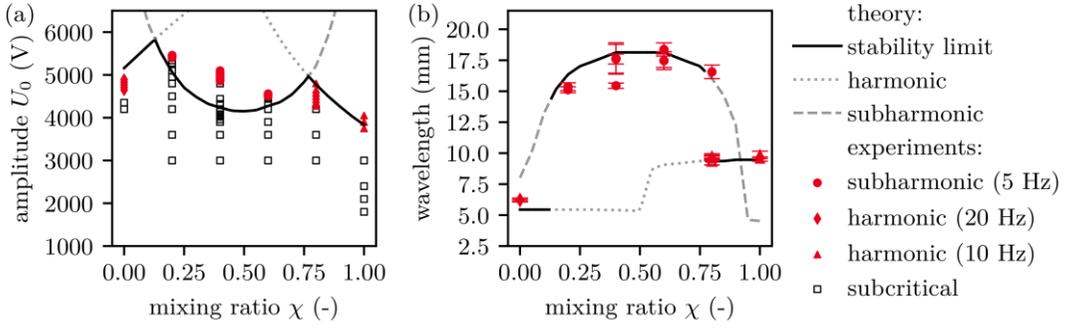

FIG. 3. Influence of the mixing ratio on the instability threshold and the pattern wavelength for AC/AC actuation. The system is excited using a voltage according to Eq. (1), with a base frequency of $f = \omega/2\pi = 10\,Hz$. (a) Experimentally obtained stability map with theoretical predictions shown as lines. (b) Dominant pattern wavelength of the Faraday waves. The stability limit indicates the response mode with the lowest critical voltage. Each data point corresponds to one critical voltage in (a). Error bars represent the standard deviation of the wavelength determined within one experiment.

Close to the intersection points, the onset voltage changes continuously with variations in $\chi$, whereas the dominant wavelengths of the Faraday waves changes discontinuously. As is shown in Fig. 3(b), the transition from the harmonic to the subharmonic branch is accompanied by a jump of the dominant wavelength, from approximately $6.2\,mm$ at $\chi = 0$ to $15.2\,mm$ at $\chi = 0.2$. Within the subharmonic branch, the wavelengths vary between $15.1\,mm$ and $18.3\,mm$. At the second intersection point, the wavelength is reduced to approximately $9.6\,mm$. Thus, changing the mixing ratio $\chi$ allows to tune the response of the interface between three different oscillation frequencies, with three associated wavelength ranges.

At this point, it is instructive to discuss the differences between electrostatic and mechanical actuation. Both actuation schemes have a different response when the single-frequency limit of Eq. (1) is considered. For electrostatic forcing, the Faraday waves oscillate isochronously with the forcing frequency $\omega$ ($2\omega$) at $\chi = 1$ ($\chi = 0$). The subharmonic response with frequency $\omega/2$ is present only for mixing ratios different from $\chi = 0,1$. Thus, we observe two bicritical points. In contrast, for most mechanical forcing conditions with a single forcing frequency $\omega$ ($2\omega$), the Faraday waves oscillate subharmonically with the frequency $\omega/2$ ($\omega$). When varying the mixing ratio $\chi$ for mechanical forcing with two frequency components $\omega$ and $2\omega$, the waves change their oscillatory frequency once, leading to only one bicritical point.

It is worth noting that close to the intersection point $\chi_2$, superpositions of the subharmonic and the harmonic response were visible. Corresponding to that, in Fig. 2 we display the temporal distribution of the wavelengths present in the system during a time interval of $200\,ms$. Here, the interface shows simultaneous harmonic and subharmonic oscillations, which we can identify from the period of oscillation of each component. When the amplitude of a specific response pattern is small, the corresponding wavelength is not detected. Since small amplitudes occur twice during one period of oscillation, we can estimate the oscillation period of the different wavelength components from these instances without detected wavelengths. As is visible in Fig. 2, the instability component with a wavelength around $16.6\,mm$ has an oscillation period of $200\,ms$ and is thus subharmonic, while the instability component with a wavelength around 9.3 mm has an oscillation period of 100 ms and is thus harmonic. Both response wavelengths correspond to the theoretically expected wavelengths. Thus, tuning the mixing ratio close to the intersection points $\chi_1, \chi_2$ also allows creating superposed oscillations. It is worth noting that for mechanical multi-frequency actuation, the instability modes close to bicritical points have shown interesting spatial distributions such as quasi-crystalline patterns with symmetry in the Fourier space, and thus have been studied extensively. Here, we solely focus on the resulting dominant wavelengths of the system obtainable by linear stability analysis, without characterizing the spatial distribution of patterns resulting from nonlinear interactions [24,43,44]. However, we expect very similar effects to be observable for electrostatic forcing.

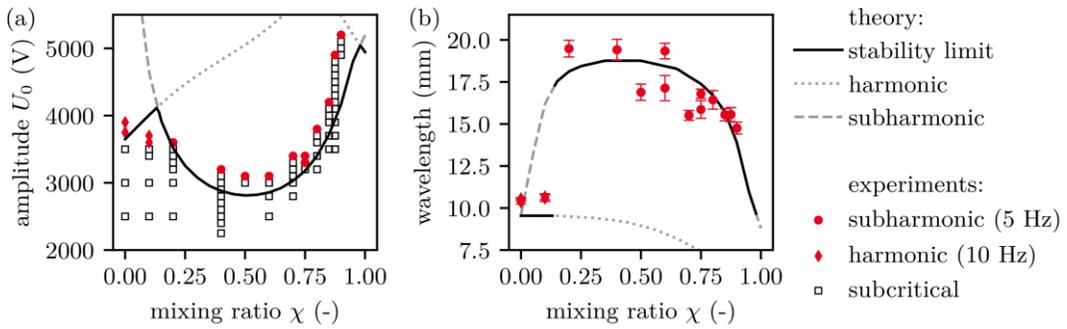

FIG. 4. Influence of the mixing ratio on the instability threshold and the pattern wavelength for AC/DC actuation. The system is excited using a voltage according to Eq. (2), with a base frequency of $f = \omega/2\pi = 10\ Hz$. (a) Experimentally obtained stability map with theoretical predictions shown as lines. (b) Dominant pattern wavelength of the Faraday waves. The stability limit indicates the response mode with the lowest critical voltage. Each data point corresponds to one critical voltage in (a). Error bars represent the standard deviation of the wavelength determined within one experiment.

*AC/DC actuation.* We study the instability induced by the superposition of an AC and a DC signal of the form of Eq. (2), where the base frequency is $f = \omega/2\pi = 10$ Hz. Adding a constant voltage has two effects on the Maxwell stress at the interface, which scales as the square of the electric field: First, the forcing now exhibits frequency components with $f$ and $2f$, which is of importance for the edge waves. Second, it adds a constant component to the Maxwell stress, which is directed in the positive $z$-direction, opposing gravity. The ability to apply a constant forcing is a qualitative difference to the mechanically forced Faraday instability, where such options are limited. It offers an additional degree of control over the system response. The Faraday instability in response to an AC signal mixed with a constant DC offset was studied by Ward et al. [41] for one test case. However, since the AC amplitude was varied between experiments, the mixing ratio χ varied as well. Here, we focus on the mixing ratio as a means to control the Faraday waves.

Figure 4(a) shows the stability map for the AC/DC driving in comparison to theoretical predictions. For small χ, corresponding to a relatively small DC component, the interface oscillates harmonically, in line with previous experimental investigations [41,42]. At a mixing ratio of approximately $\chi_{DC} \approx 0.14$, a crossover occurs, and the critical voltage of the subharmonic response becomes smaller than that of the harmonic response. Above $\chi_{DC}$ the interface oscillates subharmonically with 5 Hz. At high mixing ratios ($\chi > 0.9$), we were unable to observe Faraday waves, as instabilities lead to contact between the conducting liquid and the upper electrode without the excitation of Faraday waves. For the range of accessible χ, the experimentally measured critical amplitude agrees reasonably well with the theoretical predictions.

Figure 4(b) displays the dominant wavelength of the system, which show a discontinuous behavior at the intersection point $\chi_{DC}$, changing from approximately 10.5 mm at $\chi = 0.1$ to 19.5 mm at $\chi = 0.2$. For intermediate mixing ratios, the response wavelength is approximately constant, until it decreases at high $\chi$. This behavior is interesting, as it allows tuning the Faraday wave structure continuously by adding a constant offset to the driving voltage.

*Summary and conclusions.* In this letter, we studied the electrostatic Faraday instability induced by a superposition of an AC signal with either a second AC signal or a DC signal. An algorithm based on light refraction at the interface allows obtaining the interface deflection. We measure both the critical voltage and the resulting pattern wavelengths as a function of the mixing ratio, which denotes the relative importance of each component. For the AC/AC actuation, the liquid-liquid interface can oscillate with three different frequencies, depending on $\chi$. The obtained wavelength is discontinuous with $\chi$, showing a jump when the interface transitions from subharmonic to harmonic oscillations. Also, close to the crossover points, subharmonic and harmonic responses can superpose. For AC/DC actuation, the interface starts oscillating subharmonically above a critical mixing ratio. The mixing ratio allows tuning the wavelength continuously in a finite range, thus providing a degree of control over the wave patterns that is very difficult to achieve by mechanical forcing. Our results both validate theoretical predictions for a broad class of forcing functions [40] and highlight the unique possibilities electrostatic forcing provides in Faraday instabilities.

We thank J. Keller and J. Bültemann for designing an initial version of the set-up and the technical support. We acknowledge funding by the Deutsche Forschungsgemeinschaft (S.D., M.H., S.H., Grant No. HA 2696/50-1) and the Council of Scientific and Industrial Research (A.B., grant no. 22(0843)/20/EMR-II).

―――――――――――――――――――――――――――